\documentclass[journal,article,accept,moreauthors,10pt,a4paper]{Definitions/mdpi} 

\firstpage{1} 
\pubvolume{xx}
\issuenum{1}
\articlenumber{5}
\pubyear{2018}
\copyrightyear{2018}
\history{Received: date; Accepted: date; Published: date}
\usepackage{color}
\usepackage{amsfonts}
\usepackage{amssymb}
\usepackage{amsmath}
\usepackage{setspace} 
\usepackage{bm} 
\usepackage{mathrsfs}
\usepackage[utf8]{inputenc}
\usepackage[normalem]{ulem}  % \sout{old text} for strikeout

\newcommand{\tr}{\mathop{\mathrm{tr}}}
\newcommand{\Tr}{\mathop{\mathrm{Tr}}}

\newcommand{\hH}{\hat{H}}
\newcommand{\hN}{\hat{N}}
\newcommand{\hK}{\hat{K}}
\newcommand{\hT}{\hat{T}}

\newcommand{\ptc}[1]{{\bar{#1}}}
\newcommand{\hSigma}{\hat{\Sigma}}
\newcommand{\hS}{\hat{S}}
\newcommand{\hJ}{\hat{J}}

\newcommand{\hcurrent}{\hat{\mathcal{J}}}

\newcommand{\hO}{\hat{\mathcal{O}}}
\newcommand{\hs}{\hat{s}}

\newcommand{\lie}{\mathsterling}

\newcommand{\pt}{t}

\newcommand\hLcal{\hat{\mathcal{L}}}

\newcommand\Ncal{\mathcal{N}}

\newcommand\Scal{\mathcal{S}}

\newcommand\calG{\mathcal{G}}
\newcommand\Dcal{\mathcal{D}}

\newcommand\Acal{\mathcal{A}}

\newcommand\Fcal{\mathcal{F}}

\newcommand{\average}[1]{\langle#1\rangle}
\newcommand{\averageLG}[1]{\langle#1\rangle^\text{LG}}
\newcommand{\averageG}[1]{\langle#1\rangle_\text{eq}}
\newcommand{\AverageLG}[1]{\left\langle#1\right\rangle^\text{LG}}

\newcommand{\la}[1]{\overleftarrow{#1}}
\newcommand{\ra}[1]{\overrightarrow{#1}}

\newcommand{\with}{\quad\mathrm{with}\quad}
\newcommand{\HS}{\mathrm{hs}}
\newcommand{\con}{\mathrm{con}}
\newcommand{\ano}{\mathrm{ano}}

\newcommand{\bx}{\bm{x}}

\newcommand{\bq}{\bm{q}}
\newcommand{\hrhoLG}{\hat{\rho}_{\mathrm{LG}}}

\newcommand{\tilF}{\widetilde{F}}

\newcommand{\tile}{\widetilde{e}}
\newcommand{\tilA}{\widetilde{A}}
\newcommand{\tilD}{\widetilde{D}}
\newcommand{\tilScal}{\widetilde{\mathcal{S}}}
\newcommand{\tilAcal}{\widetilde{\mathcal{A}}}
\newcommand{\tilpartial}{\widetilde{\partial}}
\newcommand{\tilomega}{\widetilde{\omega}}
\newcommand{\tils}{\widetilde{s}}
\newcommand{\tilt}{\widetilde{t}}
\newcommand{\tilx}{\widetilde{x}}
\newcommand{\tilvare}{\widetilde{\varepsilon}}

\Title{
Anomaly-induced transport phenomena \\
from the imaginary-time formalism}

\Author{
Masaru Hongo $^{1,\dagger}$%\orcidA{}
, and Yoshimasa Hidaka $^{2,1}$}

\AuthorNames{Masaru Hongo and Yoshimasa Hidaka}

\address{%
$^{1}$ \quad RIKEN iTHEMS, RIKEN, Wako, Saitama 351-0198, Japan; masaru.hongo@riken.jp \\
$^{2}$ \quad Quantum Hadron Physics Laboratory, RIKEN Nishina Center, RIKEN, Wako, Saitama 351-0198, Japan; hidaka@riken.jp}

\corres{Correspondence: masaru.hongo@riken.jp; Tel.: +81-48-462-1226}

\abstract{
A derivation of the anomaly-induced transport phenomena---the 
chiral magnetic/vortical effect---is revisited
based on the imaginary-time formalism of quantum field theory.
Considering the simplest anomalous system composed of 
a single Weyl fermion, we provide two derivations: 
perturbative (one-loop) evaluation of the anomalous transport coefficient, 
and the anomaly matching for the local thermodynamic functional.
}


\begin{document}
%%%%%%%%%%%%%%%%%%%%%%%%%%%%%%%%%%%%%%%%%%
%% Only for the journal Gels: Please place the Experimental Section after the Conclusions

%%%%%%%%%%%%%%%%%%%%%%%%%%%%%%%%%%%%%%%%%%
% \setcounter{section}{-1} %% Remove this when starting to work on the template.
% \section{How to Use this Template}
% The template details the sections that can be used in a manuscript. Note that the order and names of article sections may differ from the requirements of the journal (e.g., the positioning of the Materials and Methods section). Please check the instructions for authors page of the journal to verify the correct order and names. For any questions, please contact the editorial office of the journal or support@mdpi.com. For LaTeX related questions please contact Janine Daum at latex-support@mdpi.com.
%The order of the section titles is: Introduction, Materials and Methods, Results, Discussion, Conclusions for these journals: aerospace,algorithms,antibodies,antioxidants,atmosphere,axioms,biomedicines,carbon,crystals,designs,diagnostics,environments,fermentation,fluids,forests,fractalfract,informatics,information,inventions,jfmk,jrfm,lubricants,neonatalscreening,neuroglia,particles,pharmaceutics,polymers,processes,technologies,viruses,vision

\section{Introduction}
Quantum anomaly is one of the most fundamental properties of quantum systems, 
which keeps staying in the low-energy regime once it appears in an underlying UV theory~\cite{Bertlmann,Fujikawa-Suzuki}.
As a consequence, the low-energy dynamics is strongly influenced by 
the existence of the quantum anomaly. 
A well-known example is the chiral anomaly in QCD,
which gives rise to the Wess-Zumino term in the low-energy effective theory 
of QCD (the chiral perturbation theory) describing the neutral pion decay 
into two photons ($\pi^0 \to \gamma \gamma$)~\cite{Fukuda-Miyamoto,Adler:1969gk,Bell:1969ts}.
The notion of anomaly can be  generalized to discrete symmetries
of systems such as time-reversal symmetry. 
The anomaly matching argument~\cite{tHooft:1979rat,Frishman:1980dq} is actively applied to restrict the possible nontrivial ground states (See Refs.~\cite{Wen:2013oza,Tachikawa:2016cha,Gaiotto:2017yup,Tanizaki:2017bam,Shimizu:2017asf,Tanizaki:2017qhf,Tanizaki:2017mtm,Sulejmanpasic:2018upi,Yao:2018kel,Tanizaki:2018xto,Tanizaki:2018wtg,Yonekura:2019vyz} for recent applications).

\begin{figure}[t]
\centering
\includegraphics[width=0.6\linewidth]{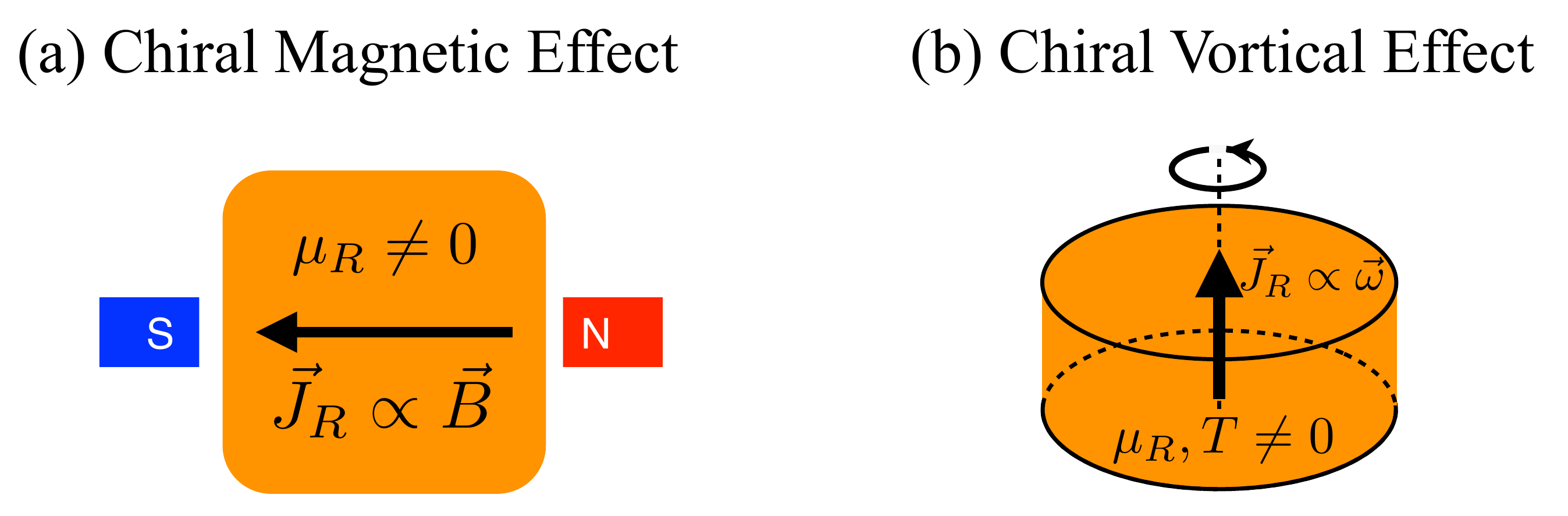}
\caption{
 The schematic picture of the anomaly-induced transport phenomena:
 (\textbf{a}) Chiral magnetic effect. (\textbf{b}) Chiral vortical effect.
  }
 \label{fig:CME-CVE}
\end{figure}

It has been recently noticed that quantum anomaly
also appears even in the effective theory describing the real-time dynamics of 
nonequilibrium systems, e.g., hydrodynamics and the kinetic theory,
and it affects the macroscopic transport properties in the 
hydrodynamic regime~\cite{Fukushima:2008xe,Erdmenger:2008rm,Banerjee:2008th,Torabian:2009qk,Son:2009tf,Amado:2011zx,Landsteiner:2011cp,Gao:2012ix,Son:2012wh,Banerjee:2012iz,Jensen:2012jh,Jensen:2012jy,Banerjee:2012cr,Stephanov:2012ki,Landsteiner:2012kd,Jensen:2012kj,Son:2012zy,Chen:2012ca,Jensen:2013vta,Jensen:2013kka,Jensen:2013rga,Manuel:2013zaa,Chen:2014cla,Haehl:2015pja,Chen:2015gta,Golkar:2015oxw,Chowdhury:2016cmh,Landsteiner:2016led,Gorbar:2016ygi,Hidaka:2016yjf,Hidaka:2017auj,Mueller:2017lzw,Mueller:2017arw,Glorioso:2017lcn,Hidaka:2018ekt,Carignano:2018gqt,Dayi:2018xdy,Liu:2018xip,Mueller:2019gjj} (See also pioneering works by Vilenkin~\cite{Vilenkin:1979ui,Vilenkin:1980fu}).
For example, the simplest anomalous system composed of a single 
right-handed Weyl fermion coupled to a background electromagnetic field 
shows interesting transport 
.
When this system is put into an environment with a temperature $T$ and 
a chemical potential $\mu_R$, the chiral anomaly induces the dissipationless 
current along the magnetic field $B^i$ given by
\begin{equation}
 \average{\hJ_{R}^i}_{{\ano}} = \sigma_B B^i + \sigma_\omega \omega^i
  \with 
  \sigma_B = \frac{\mu_R}{4\pi^2} , \quad
  \sigma_\omega = \frac{\mu_R^2}{4\pi^2} + \frac{T^2}{12} ,
  \label{eq:CME-CVE}
\end{equation}
where $\average{\hJ_{R}^\mu}_{{\ano}}$ denotes 
the anomalous part of the expectation value of 
the right-handed current, and $\sigma_B$ ($\sigma_\omega$) 
is regarded as the chiral magnetic (vortical) conductivity.
The first and second terms in Eq.~\eqref{eq:CME-CVE} are called 
the chiral magnetic effect (CME) and chiral vortical effect (CVE), respectively
(See Fig. \ref{fig:CME-CVE}).
It is worth pointing out that even in the weak coupling limit, $\sigma_B$ and 
$\sigma_\omega$ do not diverge unlike the usual conductivity because 
their existence  is protected by the quantum anomaly. 

These anomalous transports are believed to be universally present 
when the system under consideration contains the chiral anomaly. 
For example, they are expected to take place 
in the quark-gluon plasma created in high-energy heavy-ion collisions~\cite{Kharzeev:2007jp,Kharzeev:2010gd,Burnier:2011bf,Hongo:2013cqa,Yee:2013cya,Hirono:2014oda,Adamczyk:2015eqo,Yin:2015fca,Huang:2015oca,Kharzeev:2015znc,Shi:2017cpu},
astrophysical plasma including neutrino process~\cite{Charbonneau:2009ax,Grabowska:2014efa,Kaminski:2014jda,Sigl:2015xva,Yamamoto:2015gzz,Masada:2018swb}, and Weyl semimetals 
realized in condensed matter physics~\cite{Zyuzin:2012tv,Goswami:2012db,Chen:2013mea,Basar:2013iaa,Hosur:2013kxa,Landsteiner:2013sja,Chernodub:2013kya,Gorbar:2013dha,Armitage:2017cjs,Gorbar:2017lnp}.
While we have not observed clear experimental signal of the anomaly-induced transport
in the first two systems, it has been recently reported that the experimental 
signal of the CME are achieved in the Weyl semimetal~\cite{Li:2014bha,Lv:2015pya,Xu:2015cga}.

The theoretical derivation of the anomaly-induced transport phenomena has been 
remarkably developed in the past ten years, e.g., 
the direct field theoretical evaluation~\cite{Fukushima:2008xe}, 
the fluid/gravity correspondence~\cite{Banerjee:2008th,Erdmenger:2008rm,Torabian:2009qk,Amado:2011zx}, 
the phenomenological entropy-current analysis~\cite{Son:2009tf}, 
the linear response theory~\cite{Landsteiner:2011cp,Jensen:2012jy,Landsteiner:2012kd}, 
the kinetic theory~\cite{Gao:2012ix,Son:2012wh,Stephanov:2012ki,Son:2012zy,Chen:2012ca,Manuel:2013zaa,Chen:2014cla,Chen:2015gta,Gorbar:2016ygi,Hidaka:2016yjf,Hidaka:2017auj,Mueller:2017lzw,Mueller:2017arw,Hidaka:2018ekt,Carignano:2018gqt,Dayi:2018xdy,Liu:2018xip,Mueller:2019gjj},
and the hydrostatic partition function method and extensions~\cite{Banerjee:2012iz,Jensen:2012jh,Jensen:2012kj,Banerjee:2012cr,Jensen:2013vta,Jensen:2013kka,Jensen:2013rga,Haehl:2015pja,Golkar:2015oxw,Chowdhury:2016cmh,Glorioso:2017lcn}.
In this paper, we review the derivation of the anomaly-induced transport 
phenomena from the statistical mechanical viewpoint with the help of
the imaginary-time (Matsubara) formalism of quantum field theory~\cite{Matsubara,AGD,le2000thermal,KapustaGale}.
In particular, we demonstrate two derivations, which are basically 
on the same line as the last two derivations raised above.
For that purpose, we consider the simplest anomalous system composed of 
a single Weyl fermion coupled to an external electromagnetic field.
Although most results given in this paper has been already known,
we gives the clear rigorous justification of the hydrostatic partition function 
method for the anomalous system based on the statistical ensemble 
describing systems in general local thermal equilibrium.
This shows that the hydrostatic partition function method is indeed 
not restricted to the real hydrostatic situation,
but applicable to systems in general local thermal equilibrium.

The paper is organized as follows: 
In Sec.~\ref{sec:Setup}, we review the basic setup and formulation
including the Zubarev's nonequilibrium statistical operator methods~\cite{Zubarev:1979,Zubarev1,Zubarev2} 
(See also Refs.~\cite{Becattini:2014yxa,Hayata:2015lga,Hongo:2016mqm,Hongo2019} for a recent sophisticated revival of a similar idea).
In Sec.~\ref{sec:Oneloop},
we then provide the perturbative evaluation of the chiral magnetic/vortical
conductivity with the help of the (equilibrium) linear response theory, 
from which we can read off the constitutive relation for the anomalous current.
In Sec.~\ref{sec:Matching}, we give another nonperturbative derivation 
based on the anomaly matching for the local thermodynamic functional.
Sec.~\ref{sec:Summary} is devoted to the summary and discussion.

\section{Preliminaries for the anomaly-induced transport phenomena}
\label{sec:Setup}

In this section, we briefly summarize the formulation to derive the 
anomaly-induced transport phenomena based on the imaginary-time 
formalism~\cite{Zubarev:1979,Zubarev1,Zubarev2,Becattini:2014yxa,Hayata:2015lga,Hongo:2016mqm,Hongo2019}.

\subsection{Anomalous (non-)conservation laws for a single Weyl fermion}
Let us consider the system consisting of a right-handed Weyl fermions 
$\xi$ under an external $U(1)$ gauge field $A_\mu$ in a $(3+1)$ dimensional
curved spacetime, whose action has the form:  
\begin{equation}
 \Scal [\xi,\xi^\dag;A_\mu,e_\mu^{~a}] =  
  \int d^4 x e
  \left[\frac{i}{2} \xi^\dag 
  \left( e_a^{~\mu} \sigma^a \ra{D}_\mu  
    - \la{D}_\mu \sigma^a e_a^{~\mu}  \right) \xi 
	 \right]
  \with 
  e \equiv \mathrm{det} (e_\mu^{~a}),
  \label{eq:Action}
\end{equation}
where we introduced $\sigma^a = ( 1, \sigma^i)$ with the Pauli matrices 
$\sigma^i~(i=1,2,3)$. 
Here $e_\mu^{~a} (e^\mu_{~a})$ denotes (inverse) vierbein satisfying 
$g_{\mu\nu} = e_\mu^{~a} e_\nu^{~b} \eta_{ab},~
\eta^{ab} = e_\mu^{~a} e_\nu^{~b} g^{\mu\nu}$ 
with the spacetime curved metric $g_{\mu\nu}$ and Minkowski metric 
$\eta_{ab} = \mathrm{diag} (-1,+1,+1,+1)$.
The left and right covariant derivatives are defined as 
\begin{equation}
 \ra{D}_\mu \xi 
  \equiv \partial_\mu \xi - i (\Acal_\mu + A_\mu) \xi, 
  \quad 
 \xi ^{\dag} \la{D}_\mu 
  \equiv \partial_\mu \xi^\dag + i \xi^\dag (\Acal_\mu + A_\mu)
  \with 
  \Acal_\mu \equiv \frac{1}{2} \omega_\mu^{~ab} \Sigma_{ab},
\end{equation}
where we introduced
$\Sigma_{ab} \equiv i (\sigma^a \bar{\sigma}^b - \sigma^b \bar{\sigma}^a)/4$
with $\bar{\sigma}^a \equiv (-1, \sigma^i)$, which satisfies
$\sigma^a \bar{\sigma}^b + \sigma^b \bar{\sigma}^a = 2 \eta^{ab}$.
Furthermore, employing the torsionless condition, 
we can express the spin connection $\omega_\mu^{~ab} = - \omega_\mu^{~ba}$ 
as
\begin{equation}
 \omega_\mu^{~ab} 
  \equiv \frac{1}{2} e^{a\nu}e^{b\rho} 
  (C_{\nu\rho\mu} - C_{\rho\nu\mu} - C_{\mu\nu\rho}) 
  \with
  C_{\mu\nu\rho} \equiv
  e_\mu^{~c} (\partial_\nu e_{\rho c} - \partial_\rho e_{\nu c}).
  \label{eq:SpinC}
\end{equation}
Although the classical action \eqref{eq:Action} is invariant 
under a set of infinitesimal diffeomorphism, local Lorentz, 
and $U(1)$ gauge transformations with parameters 
$\chi  \equiv \{ \zeta^\mu,\, \alpha_{ab},\, \theta \}$:
\begin{equation}
 \begin{cases}
  \delta_\chi e_\mu^{~a} 
  = \zeta^\nu \nabla_\nu e_\mu^{~a}
   +  e_\nu^{~a} \nabla_\mu \zeta^{\nu} 
  + \alpha^a_{~b} e_\mu^{~b},
  \\
  \delta_\chi A_\mu
  = \zeta^\nu \nabla_\nu A_\mu + A_\nu \nabla_\mu \zeta^{\nu} 
  + \partial_\mu \theta,
  \vspace{2pt} \\
  \delta_\chi \xi
  = \zeta^\nu \partial_\nu \xi - \dfrac{i}{2} \alpha^{ab} \Sigma_{ab} \xi 
  + i \theta \xi,
 \end{cases}
 \label{eq:GaugeVar1}
\end{equation}   
we encounter with the quantum anomaly attached to the Weyl fermion.
As a consequence, the anomalous Ward-Takahashi identities results in the 
following operator identities corresponding to the (non-)conservation 
laws:
\begin{equation}
 \begin{cases}
  \nabla_\mu \hT^{\mu}_{~\nu} = F_{\nu\mu} \hJ^\mu,
  \vspace{2pt} \\
  \nabla_\mu \hJ^\mu 
   = -\dfrac{1}{8}C \varepsilon^{\mu\nu\rho\sigma} F_{\mu\nu} F_{\rho\sigma}
  - \lambda \varepsilon^{\mu\nu\rho\sigma} 
  R^{\alpha}_{~\beta\mu\nu} R^{\beta}_{~\alpha\rho\sigma},
 \end{cases}
  \with 
  \hT^{ab} - \hT^{ba} = 0 ,
  \label{eq:Conservation}
\end{equation}
where we introduced the energy-momentum tensor $\hT^\mu_{~\nu}$, 
$U(1)$ covariant charge current $\hJ^\mu$, 
a field strength tensor for the background electromagnetic field 
$F_{\mu\nu} \equiv \partial_\mu A_\nu - \partial_\nu A_\mu$, and
the Riemann curvature tensor $R^\mu_{~\nu\rho\sigma}$ with 
the totally antisymmetric tensor $\varepsilon^{\mu\nu\rho\sigma}$ 
satisfying $\varepsilon^{0123} = 1/e$.
For notational simplicity, we drop the subscript $R$ for the 
$U(1)$ current.
Here $C =1/(4\pi^2)$ and $\lambda =1/(768\pi^2)$ 
denote the anomaly coefficients coming from 
gauge and gravitational sectors, respectively.
Since $\lambda \varepsilon^{\mu\nu\rho\sigma} R^{\alpha}_{~\beta\mu\nu} R^{\beta}_{~\alpha\rho\sigma}$ contains four derivatives, it does not contribute to the first order hydrodynamics that we are interested in.
  Therefore, we will omit the gravitational part in the following discussion. 
  Note that while the gauge and diffeomorphism invariance provides 
two (non-)conservation laws, the local Lorentz invariance results in 
the symmetric property of the energy-momentum tensor operator.
It is worth emphasizing that $\hJ^\mu$ in Eq.~\eqref{eq:Conservation} 
is the covariant current which can be related to the consistent current 
$\hJ_{\con}^\mu$ by 
\begin{equation}
 \hJ^\mu 
  = \hJ_{\con}^\mu
  - \frac{1}{6} C \varepsilon^{\mu\nu\rho\sigma} A_\nu F_{\rho\sigma}.
\end{equation}
An analogue of this relation in local thermal equilibrium will appear 
in Sec.~\ref{sec:Matching}, and it plays an important role to see
how the anomaly matching is realized for the local thermodynamic functional.

\subsection{Zubarev's formula: Decomposing dissipative and nondissipative transport}
We then briefly review the Zubarev's nonequilibrium statistical operator method 
from the modern viewpoint (See e.g., Refs.~\cite{Zubarev:1979,Zubarev1,Zubarev2,Becattini:2014yxa,Hayata:2015lga,Hongo:2016mqm,Hongo2019} for recent discussions) 
and specify from where the anomaly-induced transport arises.
Assuming that the system is initially in local thermal equilibrium, 
the Zubarev's formula provides us the expectation values of  
conserved current operators 
$\hcurrent^\mu_{~a} \equiv \{\hT^\mu_{~\nu},\,\hJ^\mu\}$ over the 
initial density operator in the following compact form:
\begin{equation}
 \average{\hcurrent^\mu_{~a}(t,\bx)}
  = \averageLG{\hcurrent^\mu_{~a} (t,\bx)}_{t} + 
  L^{\mu\nu}_{ab} (t,\bx) \nabla_\nu \lambda^b (t,\bx)
  + O \big( (\nabla \lambda)^2 \big),
  \label{eq:Zubarev}
\end{equation}
where we introduced the intensive local thermodynamic parameters
$\lambda^a \equiv \{\beta^\mu,\,\nu\}$, which are related to
the local fluid temperature $T = 1/\beta$, four-velocity $u^\mu$, 
and the chemical potential $\mu$ through
$\beta^\mu \equiv \beta u^\mu,~\nu \equiv \beta \mu$.
We also defined the average over the local Gibbs distribution 
$\hrhoLG[\lambda;t]$, which describes systems in local thermal equilibrium,
for an arbitrary operator $\hO$ as
\begin{equation}
 \averageLG{\hO}_{t}
  \equiv \Tr 
  \left( \hrhoLG [\lambda;t] \hO \right)
  \with 
  \hrhoLG [\lambda;t]
  \equiv \exp \left[ - \hS [\lambda;t] \right], 
  \quad 
  \hS [\lambda;t] = \hK[\lambda;t] + \Psi [\lambda;t],
\end{equation}
where the entropy operator $\hS[\lambda;t]$ is composed of 
the part including operators $\hcurrent^\mu_{~a}$ and 
normalization part for the density operator:
\begin{align}
 \hK [\lambda;t] 
 &\equiv - \int d \Sigma_{t \mu} 
  \left[ \beta^\nu (t,\bx) \hT^\mu_{~\nu} (t,\bx)
   + \nu (t,\bx) \hJ^\mu (t,\bx) \right],
 \label{eq:hatK}
 \\
 \Psi [\lambda;t] 
 &\equiv \log \Tr \exp \left[ - \hK [\lambda;t] \right].
 \label{eq:MP}
\end{align}
We here employed the fully covariant notion by introducing the constant 
time (spacelike) hypersurface defined by its perpendicular surface vector 
$d\Sigma_{t\mu} \equiv -d^3 x \sqrt{\gamma} n_\mu$.
Choosing a certain globally defined time-coordinate function $\bar{t}(x)$,
the unit normal vector $n_\mu$ can be expressed as
\begin{equation}
 n_\mu (x) = - N (x) \partial_\mu \bar{t}(x)
  \with
  N (x) \equiv 
  \left( -\partial^\mu \bar{t} (x) \partial_\mu \bar{t} (x) \right)^{-1/2},
\end{equation}
where $N(x)$ is a so-called Lapse function.
In addition, introducing the spatial coordinate on the $\bar{\bx}$, 
we have the induced metric $\gamma_{\mu\nu} = g_{\mu\nu} + n_\mu n_\nu$
whose spatial part gives $\gamma \equiv \mathrm{det} \gamma_{\ptc{i}\ptc{j}}$ 
(See e.g.,~Refs.~\cite{Hayata:2015lga,Hongo:2016mqm} for a detailed geometric setup). 
The introduction of the covariantized notion looks a little bit complicated, 
but one can always take the flat limit by setting 
$\big( \bar{t}(x), \bar{\bx} (x)\big) = (t,\bx)$, 
which results in e.\,g. 
$d\Sigma_{t\mu}|_{\mathrm{flat}} =  d^3 x \delta_\mu^0$.
Although it might be desirable to distinguish two coordinate systems 
defined by $(t,\bx)$ and $(\bar{t},\bar{\bx})$, we will basically omit 
overline for the later one for notational simplicity 
since only $(\bar{t},\bar{\bx})$-coordinate system is mainly used.
The normalization part $\Psi [\lambda;t]$ is the local thermodynamic functional 
called the Massieu-Planck functional, and plays a central role in 
Sec.~\ref{sec:Matching}.

The crucial point here is that, by construction, 
we can identify the first term in the right-hand-side of 
Eq.~\eqref{eq:Zubarev} as the nondissipative transport taking place in locally 
thermalized system, whereas the second term as the dissipative correction
coming from the deviation from local thermal equilibrium.
In other words, the formula \eqref{eq:Zubarev} gives a way to decompose the 
non-dissipative and dissipative transport at least in the leading-order 
derivative expansion.
The second term is proportional to the (local) thermodynamic forces 
$\nabla_\nu \lambda^b$, and coefficients in front of them are indeed specified 
as transport coefficients such as the bulk/shear viscosity, and conductivity.
They are expressed by the two-point (Kubo) correlation function, which 
is nothing but the Green-Kubo formula for the transport coefficient~\cite{Zubarev:1979,Zubarev1,Zubarev2,Becattini:2014yxa,Hayata:2015lga,Hongo:2016mqm,Hongo2019}.
On the other hand, nondissipative part is often assumed to be simply given
by the usual constitutive relation for a perfect fluid.
This is the case for parity-invariant systems, 
since the nondissipative derivative corrections are accompanied with 
higher-order derivatives for parity-invariant systems.
Nevertheless, if we consider a system without parity symmetry---like 
the Weyl fermion system given in Eq.~\eqref{eq:Action}---we generally 
encounter with first-order nondissipative derivative corrections 
in $\averageLG{\hcurrent^\mu_{~a}(t,\bx)}_t$.
This is the origin of the anomaly-induced transport, and we will 
focus on how we can evaluate $\averageLG{\hcurrent^\mu_{~a}(t,\bx)}_t$ 
in the remaining part of this paper.

Before closing this section, we put a short comment on the absence of the 
anomalous contribution to the entropy production.
To see this, using the conservation laws \eqref{eq:Conservation},
we express the entropy production operator 
$\hSigma [t,t_0;\lambda] \equiv \hS [\lambda;t] - \hS [\lambda;t_0]$ as 
\begin{equation}
  \hSigma [t,t_0;\lambda]   
  = \int_{t_0}^t d^4 x e \nabla_\mu \hs^\mu
  \with
  \nabla_\mu \hs^\mu \equiv
  - (\nabla_\nu\beta^\mu) \delta {\hT^{\mu}}_{~{\nu}}  
  - ( \nabla_\mu \nu + \beta^\nu  F_{\mu\nu}) \delta \hJ^{\mu} ,
\end{equation}
where we defined the local entropy production rate $\nabla_\mu \hs^\mu$ 
with $\delta \hO (t) \equiv \hO (t) - \averageLG{\hO(t)}_t$.
We thus find that the local equilibrium part of the constitutive relation 
$\averageLG{\hcurrent^\mu_{~a}}_t$ which also contains 
the anomaly-induced transport as first-order derivative corrections,
does not contribute to the local entropy production.
This is perfectly consistent with the phenomenological derivation of 
the anomaly-induced transport based on the entropy-current analysis given 
in Ref.~\cite{Son:2009tf}.

\section{Perturbative evaluation of anomalous transport coefficients}
\label{sec:Oneloop}

In this section, we provide a simple perturbative derivation of
the anomaly-induced transport given in Eq.~\eqref{eq:CME-CVE}, 
and calculate anomalous transport coefficients $\sigma_B$ and $\sigma_\omega$ 
at the one-loop level.

\subsection{Derivative expansion of the local Gibbs distribution}
\label{sec:SettingPerturbation}

First of all, we note that the local equilibrium part of the 
constitutive relation, or $\averageLG{\hcurrent^\mu_{~a}(t,\bx)}_t$, 
is a functional of local thermodynamic parameters 
$\lambda^a = \{\beta^\mu,\,\nu\}$ and external fields
$j \equiv \{A_\mu,\,e_\mu^{~a}\}$ at a fixed constant time $t$
since the local Gibbs distribution $\hrhoLG [\lambda;t]$ 
depends on the configuration of them.
Thus, $\averageLG{\hcurrent^\mu_{~a}(t,\bx)}_t$ inherently contains 
the derivative correction coming from the local Gibbs distribution itself.

Suppose that our system is described by the local Gibbs distribution 
slightly deviated from the global equilibrium (Gibbs) distribution
only with  the magnetic field and fluid vorticity.
We also turn off the external fields and take the flat limit.
In that situation, approximating the fluid velocity and the magnetic field as 
\begin{equation}
 \begin{cases}
  u_j (x)
  =  (x^i-x_0^i) \partial_i u_j |_{x=x_0} 
  = (x^i - x_0^i) \epsilon_{ijk} \omega^k,
  \vspace{2pt} \\
  A_j (x)
  = (x^i - x_0^i) \partial_i A_j |_{x=x_0}
  = \dfrac{1}{2} (x^i - x_0^i) \epsilon_{ijk} B^k ,
 \end{cases}
\end{equation}
we can expand the local Gibbs distribution 
on the top of the global Gibbs distribution as 
\begin{equation}
 \hrhoLG [\lambda;t]
  = \frac{1}{Z} e^{-\beta (\hH - \mu \hN)} 
  \left[ 1 + T_\tau \int_0^\beta d\tau 
   \Delta \hS (t-i\tau)
  \right]
  \with
  \Delta \hS 
  \equiv  
  \frac{1}{2} 
   \int d^3 x \epsilon_{ijk} (x^i - x_0^i)
   \left(
    \hJ^j B^k + 2 \hT^{0j} \omega^k
   \right),
\end{equation}   
where we defined 
$\hO (t-i\tau) \equiv 
e^{\tau (\hH - \mu \hN)} \hO (t) e^{-\tau (\hH - \mu\hN)}$.
Here $Z \equiv \Tr e^{-\beta (\hH - \mu \hN)}$ denotes the partition function 
for the globally thermalized system, and we use 
$\averageG{\hO} \equiv \Tr (e^{-\beta (\hH - \mu \hN)} \hO)/Z$.
Then, noting that the averaged current in global thermal equilibrium 
vanishes $\averageG{\hJ^i (t,\bx_0)} = 0$, 
we can evaluate $\averageLG{\hJ^i (t,\bx_0)}_t$ as 
\begin{equation}
 \begin{split}
  \averageLG{\hJ^i (t,\bx_0)}_t 
  &= \frac{1}{2} \int_0^\beta d\tau \int d^3 x
  \epsilon_{jkl} (x^j - x_0^j) 
  \\
  &\hspace{20pt} \times 
  \left[
  \averageG{\hJ^k (t-i\tau,\bx) \hJ^i (t,\bx_0)} 
  B^l (t,\bx_0)
  + 2 \averageG{\hT^{0k} (t-i\tau,\bx) \hJ^i (t,\bx_0)} 
  \omega^l (t,\bx_0)
  \right]
  \\
  &= \frac{i}{2} 
  \epsilon_{jkl} 
  \left[
  \partial_{q_j} \Delta_{J^k J^i} (\omega_n,\bq) \big|_{\omega_n=0,\,\bq =\bm{0}}
  B^l (t,\bx_0)
  + 2 \partial_{q_j} \Delta_{T^{0k} J^i} (\omega_n,\bq)\big|_{\omega_n=0,\,\bq =\bm{0}}
  \omega^l (t,\bx_0)
  \right],
 \end{split}
 \label{eq:Linear}
\end{equation}
where we performed the Fourier transformation to proceed the second line.
It is now clear that we only need to evaluate two-point 
imaginary-time---not real-time---correlation functions,
namely $\averageG{\hJ^k (t - i\tau,\bx) \hJ^i (t,\bx_0)}$ and 
$\averageG{\hT^{0k} (t - i\tau,\bx) \hJ^i (t,\bx_0)}$, 
or their low-frequency and wave-number in the Fourier space.

\subsection{One-loop evaluation of anomalous transport coefficients}
\label{sec:TwoPoint}

We then evaluate the anomalous transport coefficients 
with the help of the Matsubara formalism.
Since we expand the local Gibbs distribution on the top of global Gibbs 
distribution, the Euclidean action $S_E[\xi,\xi^\dag,\mu]$ 
for the right-handed Weyl fermion is simply given by
\begin{equation}
 S_E [\xi,\xi^\dag]
  = - \sum_{P} \xi^\dag_a (P) 
  \left(\calG^{-1}_0 (\tilde{P})\right)_{ab} \xi_b (P) 
  \with 
  \calG^{-1}_0 (P) \equiv \sigma^\mu P_\mu  , \quad 
   \calG_0 (P)= \frac{\ptc{\sigma}^\mu P_\mu}{P^2},
   \label{eq:FreeWeyl}
\end{equation}
where $a,b(=1,2)$ denote the spinor indices, 
and $\calG_0 (P)$ the free propagator for the Weyl fermion.
We also defined $\tilde{P}_\mu \equiv \left(-i\omega_n - \mu, \bm{p \right})$ 
with the Matsubara frequency $\omega_n \equiv(2n+1)\pi T$ 
and chemical potential $\mu$.
As usual, we introduced the Fourier transformation 
\begin{equation}
  \xi(\tau,\bx)
   = T \sum_ {n} \int \frac{d^3 \bm{p}}{(2\pi)^3} e^{-i\omega_n \tau +i \bm{p} \cdot \bm{x}} \xi(\omega_n,\bm{p}) ,
\end{equation}
with the temperature $T \equiv 1/\beta$. 
Note that the argument of the propagator in Eq.~\eqref{eq:FreeWeyl} is not $P$ but $\tilde{P}$, 
and, thus, it represents the propagator fully dressed by 
the chemical potential $\mu$. 
By using these, we need to evaluate the following diagrams:
\begin{equation}
  \parbox{4.3cm}{\includegraphics[width=4.3cm]{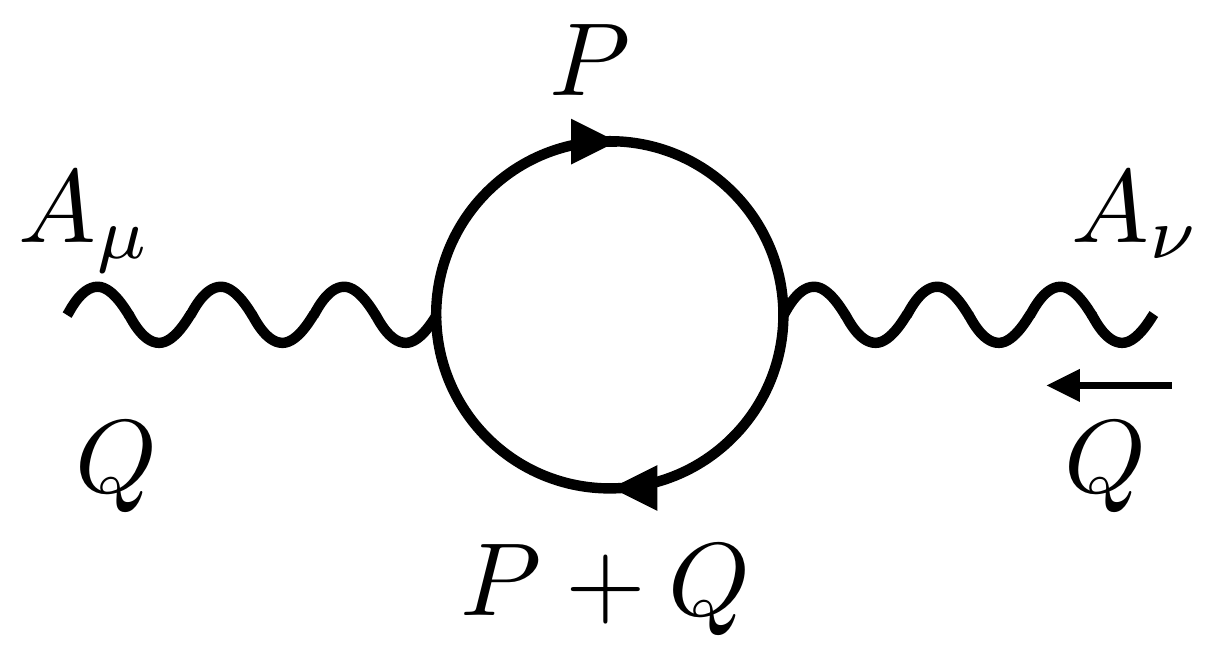} }
  \quad \mathrm{and} \quad ~
  \parbox{4.3cm}{\includegraphics[width=4.3cm]{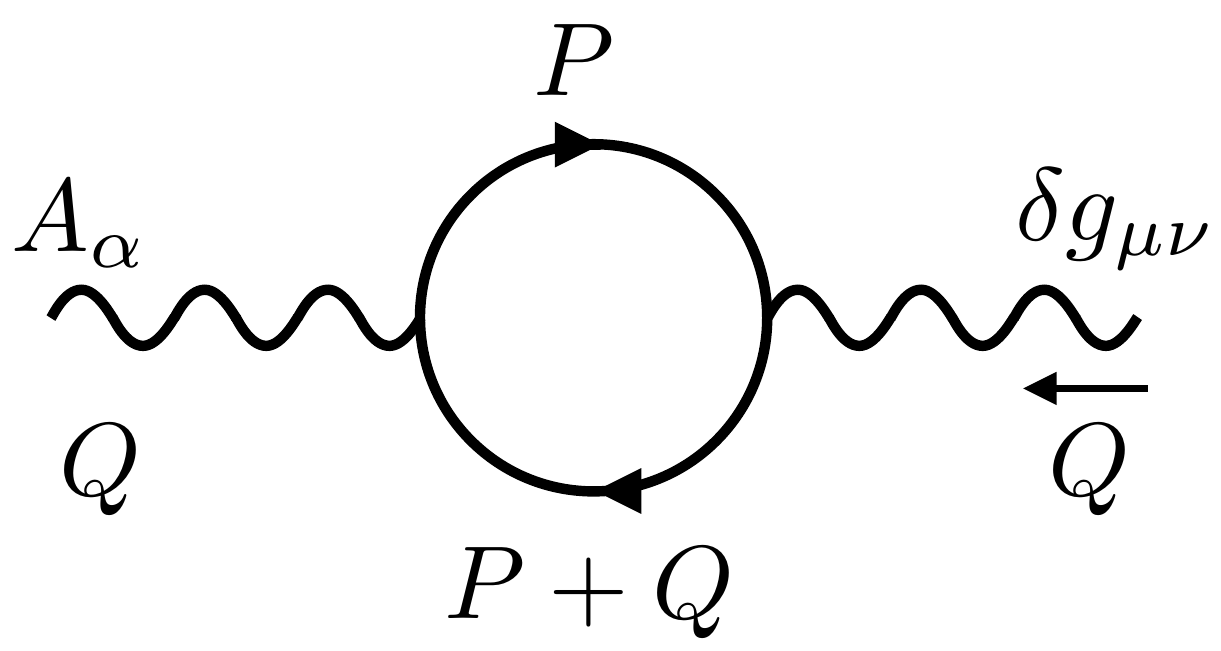}}, 
\end{equation}
where we will take the long-wave-length limit $Q \sim 0$.

First, let us evaluate the two-point current-current correlation function
given by
\begin{equation}
    \parbox{4.3cm}{\includegraphics[width=4.3cm]{fig-one-loop1.pdf} } 
     =  - T_0 \sum_n \int \frac{d^3 \bm{p}}{ (2\pi)^3} 
    \tr \left( \frac{({Q}_\sigma + \tilde{P}_\sigma ) \tilde{P}_\rho
	\ptc{\sigma}^\rho \sigma^\mu \ptc{\sigma}^ \sigma \sigma^\nu }{({Q} + \tilde{P} )^2 \tilde{P}^2 } \right ), 
\end{equation}
where we used the free propagator defined in Eq.~\eqref{eq:FreeWeyl}. 
Here ``$\tr$'' denotes the trace over the spinor indices. 
With the help of the trace formula for the Pauli matrices 
\begin{equation}
  \tr \ptc{\sigma}^\mu \sigma^\nu \ptc{\sigma}^\alpha \sigma^\beta 
  = -2i \varepsilon^{\mu\nu\alpha\beta}   + 2 \eta^{\mu\nu} \eta^{\alpha\beta} 
  - 2 \eta^{\mu\alpha}\eta^{\nu\beta} + 2 \eta^{\mu\beta} \eta^{\nu\alpha}, 
\end{equation}
we can decompose the two-point functions into the antisymmetric part and other parts. 
Since we are interested in the anomalous term which results from the antisymmetric part, we only focus on that part:
\begin{equation}
 \parbox{4.3cm}{\includegraphics[width=4.3cm]{fig-one-loop1.pdf} } 
  = - \frac{i\mu}{4\pi^2} \varepsilon^{0\mu\nu\rho}{Q}_\rho 
  + (\mathrm{symmetric~terms}) + O (Q^2), 
\end{equation}
Next, let us evaluate the two-point momentum-current correlation function.
Then, the same calculus brings about the following result
\begin{align}
 \parbox{4.3cm}{\includegraphics[width=4.3cm]{fig-one-loop2.pdf} } 
 &=- \frac{1}{4}T_0\sum_n \int \frac{d^3 \bm{p}}{ (2\pi)^3} 
 (2\tilde{P}^{\gamma}+Q^{\gamma})(\delta^{\mu}_{\beta}\delta^{\nu}_{\gamma}+\delta^{\nu}_{\beta}\delta^{\mu}_{\gamma})
   \tr \left( \frac{({Q}_\sigma + \tilde{P}_\sigma ) \tilde{P}_\rho
   \ptc{\sigma}^\rho\sigma^\beta 
 \ptc{\sigma}^ \sigma \sigma^\alpha }{({Q} + \tilde{P} )^2 \tilde{P}^2 } \right )
 \nonumber
\\
 &= i{Q}_\rho
   \left( \eta^{\nu0}\varepsilon^{\rho\mu0\alpha} +\eta^{\mu0}\varepsilon^{\rho\nu0\alpha} 
  +\delta^{\nu}_{j} \varepsilon^{\rho\mu j\alpha} +\delta^{\mu}_{j} \varepsilon^{\rho\nu j\alpha} 
   \right)\left( \frac{\mu^2}{16\pi^2} + \frac{T_0^2}{48} \right)
 \nonumber
   \\
  &\quad \hspace{130pt} + (\mathrm{symmetric~terms}) + \mathcal{O}(Q^2).
\end{align}

Putting these results all together, Eq.~\eqref{eq:Linear} results in
\begin{equation}
  \averageLG{\hJ^i (t,\bx_0)}_t 
  = 
  \frac{\mu}{4\pi^2} B^i (t,\bx_0)
  + \left( \frac{\mu^2}{4\pi^2} + \frac{T^2}{12} \right) \omega^i (t,\bx_0),
 \label{eq:CME-CVE2}
\end{equation}
which is nothing but Eq.~\eqref{eq:CME-CVE}.
To summarize the above analysis, we have derived the anomaly-induced 
transport---chiral magnetic/vortical effect---for the Weyl fermion 
by expanding the local Gibbs distribution.
This clearly shows that information on the anomaly-induced transport is fully 
contained in $\averageLG{\hcurrent^\mu_{~a}(t,\bx)}_t$.
Although we performed the direct expansion of the local Gibbs distribution
in this section, there is another way to systematically evaluate 
$\averageLG{\hcurrent^\mu_{~a}(t,\bx)}_t$ as we will see in the next section.

\section{Anomaly matching for local thermodynamic functional}
\label{sec:Matching}

In the previous section, we have explicitly shown that 
the local equilibrium part of constitutive relations
$\averageLG{\hcurrent^\mu_{~a}(t,\bx)}_t$ indeed contains 
the information on the anomaly-induced transport.
Although it is the one-loop perturbative calculation, we expect the result, or 
the value of anomalous transport coefficients, is protected by the underlying 
chiral anomaly, and remain the same even if we take into account 
the effect of interactions nonperturbatively.
In this section, we provide another way to see the anomaly-induced 
transport putting the emphasis on the nonperturbative aspect of the anomaly.
The key quantity is the local thermodynamic functional $\Psi[\lambda,j;t]$ 
already defined in Eq.~\eqref{eq:MP}.

\subsection{Basic properties of local thermodynamic functional}
\label{sec:VariationMP}
We here summarize basic properties of the 
Massieu-Planck functional $\Psi[\lambda,j;t]$:
the exact path-integral expression of $\Psi[\lambda,j;t]$ and 
resulting symmetry properties together with the variational formula.

\subsubsection{Path-integral formula and resulting symmetry}
We will first summarize the key result for the Massieu-Planck 
functional (See Refs.~\cite{Hayata:2015lga,Hongo:2016mqm} for the derivation).
Using the energy-momentum tensor operator $\hT^\mu_{~\nu}$ and 
covariant current operator $\hJ^\mu$ resulting from \eqref{eq:Action},
we can express the Massieu-Planck functional by the imaginary-time 
path integral in the same way with the usual Matsubara formalism
for global thermal equilibrium.
After a little bit tedious calculation (See Ref.~\cite{Hongo:2016mqm}), 
we eventually obtain 
\begin{equation}
 \Psi [\lambda,j;t]
  = \int \Dcal \xi \Dcal \xi^\dag 
  \exp \left( \tilScal [\xi,\xi^\dag; \tilA,\tile] \right),
  \label{eq:MPfcn}
\end{equation}
with the manifestly covariant action 
$\tilScal [\xi,\xi^\dag;\tilA_\mu,\tile_\mu^{~a}]$ given by
 \begin{equation}
 \tilScal [\xi,\xi^\dag;\tilA_\mu,\tile_\mu^{~a}] =  
  \int_0^{\beta_0} d\tau d^3 x \tile
  \left[\frac{i}{2} \xi^\dag 
   \left( \tile_a^{~\mu} \sigma^a \ra{\tilD}_\mu  
    - \la{\tilD}_\mu \sigma^a \tile_a^{~\mu} \right) \xi 
  \right]
  \with 
  \tile \equiv \mathrm{det} (\tile_\mu^{~a}).
  \label{eq:tilAction}
\end{equation}
Here we introduced the thermal (inverse) vierbein 
$\tile_\mu^{~a}\, (\tile_a^{~\mu})$ and the external $U(1)$ gauge field $\tilA_\mu$ 
in thermally emergent curved spacetime as
\begin{equation}
  \tile_{0}^{~a} = e^\sigma u^a, 
  \quad 
  \tile_{i}^{~a} = e_{i}^{~a}
  \quad
  \mathrm{and} 
  \quad
  \tilA_{0} =  e^\sigma \mu,
  \quad 
  \tilA_{i} = A_{i} ,
  \label{eq:Thermalbg}
\end{equation}
where, recalling $\beta^\mu (x) \equiv \beta (x) u^\mu (x)$ 
and $\nu (x) = \beta (x) \mu (x)$, we used
\begin{equation}
 e^{\sigma(x)} \equiv \beta(x)/\beta_0,
  \quad 
  \mu(x) \equiv \nu(x)/\beta(x),
  \quad 
  \beta (x) \equiv \sqrt{-g_{\mu\nu}(x) \beta^\mu (x)\beta^\nu (x)},
\end{equation}
with a constant reference inverse temperature $\beta_0$.
We also introduced 
$\tile \equiv \det \tile_\mu^{~a}$ and the covariant derivative 
in thermal spacetime as
\begin{equation}
 \begin{cases}
  \ra{\tilD}_\mu \xi 
  \equiv \tilpartial_\mu \xi - i (\tilAcal_\mu + \tilA_\mu) \xi, 
  \vspace{5pt} \\
  \xi^{\dag}\la{\tilD}_\mu 
  \equiv \tilpartial_\mu \xi^\dag + i \xi^\dag (\tilAcal_\mu + \tilA_\mu),
 \end{cases}
 \mathrm{with}\quad
  \tilpartial_\mu \equiv (i\partial_\tau, \partial_i),
  \quad
  \tilAcal_\mu \equiv \frac{1}{2} \tilomega_\mu^{~ab} \Sigma_{ab},
\end{equation}
where the thermal spin connection is expressed by the thermal vierbein 
$\tile_\mu^{~a}$ through the same relation in the original spacetime~\eqref{eq:SpinC}.

As is shown in these, we can say that the Massieu-Planck functional 
is expressed as the path integral in the presence of the emergent background
curved spacetime and $U(1)$ gauge field.
Note that this background structure is completely determined 
by configurations of the local thermodynamic variables $\lambda^a$ 
(and external fields $j$) on the constant time hypersurface 
in the original spacetime.
The crucial point here is that all these quantities do not depend on the 
imaginary-time coordinate $\tau$, which leads to the Kaluza-Klein gauge 
symmetry. 
To see this clearly, we express the line element 
$d\tils^2 \equiv \tile_\mu^{~a} \tile_\nu^{~b} \eta_{ab} d\tilx^\mu \otimes d\tilx^\nu$ 
and $U(1)$ gauge connection $\tilA \equiv \tilA_\mu d\tilx^\mu$ 
in thermal spacetime as
\begin{align}
 d\tils^2 
 &= - e^{2\sigma} (d\tilt + a_i dx^i)^2 + \gamma_{ij}' dx^i dx^j ,
 \label{eq:line}
 \\
 \tilA 
 &= \tilA_0 (d\tilt + a_i dx^i) + \tilA_i' dx^i,
 \label{eq:gauge}
\end{align}
with $d\tilt \equiv - i d\tau$.
Here we defined the following quantities
\begin{equation}
 a_i \equiv - e^{-\sigma} u_i, \quad
  \gamma_{ij}' \equiv 
  \gamma_{ij} {+ e^{2\sigma} a_i a_j},
  \quad
  \tilA_i' \equiv 
  = \tilA_i - \tilA_0 a_i  .
   \label{eq:KKinv}
\end{equation}
Then, in addition to the spatial diffeomorphism invariance---invariance 
under spatial coordinate transformation $\bx \to \bx'(\bx)$--- 
we now see the background \eqref{eq:line}-\eqref{eq:gauge} is 
invariant under the transformation given by
\begin{equation}
 \begin{cases}
  \tilt \to \tilt + \chi (\bx), \\
  \bx \to \bx, \\
  a_i (\bx) \to a_i (\bx) - \partial_i \chi (\bx).
 \end{cases}
\end{equation}
This is nothing but Kaluza-Klein gauge transformation, and 
$a_i$ is identified as the Kaluza-Klein gauge field.
Note that $\gamma_{ij}$ and $\tilA_i = A_i$ do transform under the 
Kaluza-Klein gauge transformation so that $\gamma_{ij}'$ and $\tilA_i'$ do not.
Therefore, it is useful to employ Kaluza-Klein gauge invariant quantities 
$\gamma_{ij}'$ and $\tilA_i'$ rather than $\gamma_{ij}$ and $\tilA_i$ 
as basic building blocks to construct the Massieu-Planck functional.
Furthermore, since the system is composed of the Weyl fermion, 
the apparent $U(1)$ gauge invariance for $\tilA_i'$ is anomalously broken.
These spatial diffeomorphism, Kaluza-Klein gauge, and 
anomalous $U(1)$ gauge symmetries provide a basic restriction to 
the Massieu-Planck functional.

\subsubsection{Variational formula in the presence of quantum anomaly}

We then provide the variational formula for the Massieu-Planck functional
$\Psi[\lambda,j;t]$, and show all information on 
$\averageLG{\hcurrent^\mu_{~a}(t,\bx)}_t$ is fully installed in it.
To show this, let us consider the variation of $\hK$ defined in 
Eq.~\eqref{eq:hatK} under the infinitesimal general coordinate 
and gauge transformation with a set of parameters 
$\zeta^\mu = \epsilon \beta^\mu$ and $\theta = \epsilon (\nu - \beta \cdot A)$.
($\epsilon$ denotes an infinitesimal constant.)
As a result of the combination of diffeomorphism and $U(1)$ 
gauge transformations, the variation of the background $U(1)$ gauge field
$\delta_\lambda A_\mu$ has the simple expression:
\begin{equation}
 \delta_\lambda A_\mu 
  = \lie_\beta A_\mu + \nabla_\mu (\nu - \beta \cdot A)
  = \nabla_\mu \nu + \beta^\nu F_{\nu\mu}.
  \label{eq:delA}
\end{equation}
The crucial point here is that 
$\hK$ remains invariant under the simultaneous transformation 
acting on both operators and external fields:
$\delta_\lambda \hK 
  \equiv 
  \delta_\lambda^{\mathrm{para}} \hK 
  + \delta_\lambda^{\mathrm{ope}} \hK  
  = 0$.
This invariance can be shown by recalling all operators in $\hK$ are $U(1)$ 
gauge invariant, and, furthermore, rewriting $\hK$ as
\begin{equation}
  \hK [\pt,\lambda^a, e_\mu^{~a},A_\mu]
  = \int d^4 x \sqrt{\gamma} \delta (\pt - \pt(x)) 
  n_\mu (x) \lambda^a (x) \hcurrent^\mu_{~a} (x),
  \label{eq:Khat1}
\end{equation}
from which we can clearly see diffeomorphism (reparametrization) invariance.
Moreover, $\delta_\lambda^{\mathrm{ope}} \hK$ 
will also trivially vanish just because 
$\delta_\lambda^{\mathrm{ope}} \hK = [i \hK, \hK] = 0$.
As a result, we have the operator identity 
$\delta_\lambda^{\mathrm{para}} \hK = 0$.

Then, let us investigate $\delta_{\lambda}^{\mathrm{para}} \hK$ in detail, 
whose explicit definition is given by
\begin{equation}
 \delta_\lambda^{\mathrm{para}} \hK
  \equiv \int d^4 x
  \left[ \frac{\delta \hK}{\delta \pt(x)} \lie_\beta \pt(x)
   + \frac{\delta \hK}{\delta \lambda^a(x)} \lie_\beta \lambda^a(x)
   + \frac{\delta \hK}{\delta e_\mu^{~a}(x)} \lie_\beta e_\mu^{~a} (x)
   + \frac{\delta \hK}{\delta A_\mu(x)} \delta_\lambda A_\mu(x) \right].
  \label{eq:delhKpara}
\end{equation}
To rewrite the first term of this equation, noting  
$\delta (\pt - \pt(x)) n_\mu = - N \delta (\pt - \pt(x)) \partial_\mu \pt
= N \partial_\mu \theta (\pt - \pt(x))$ following from 
the definition of $n_\mu$, and performing the integration by parts, 
we rewrite $\hK$ in Eq.~\eqref{eq:Khat1} as

\begin{equation}
 \begin{split}
  \hK [\pt,\lambda^a, e_\mu^{~a},A_\mu]
  &= - \int d^4 x e \theta(\pt - \pt(x))
  \nabla_\mu ( \lambda^a(x)\hcurrent_a^\mu(x) ) \\
  &= - \int d^4 x e \theta(\pt - \pt(x))
  \left( \hT^\mu_{~\nu}  \nabla_\mu \beta^\nu
  +\hJ^{\mu} (\nabla_\mu \nu + \beta^\nu F_{\nu\mu})
  - \frac{1}{8}C\nu\varepsilon^{\mu\nu\rho\sigma}
  F_{\mu\nu} F_{\rho\sigma}  \right) ,
  \label{eq:Khat2}
 \end{split}
\end{equation}
where we used $e=N \sqrt{\gamma}$ and employed the operator identity 
for current operators \eqref{eq:Conservation} to proceed the second line.
With the help of Eq.~\eqref{eq:delA} together with 
$\nabla_\mu \beta^\nu 
= e_a^{~\nu} \lie_\beta e_\mu^{~a} + \beta^\rho \omega_{\rho~\mu}^{~\nu}$ 
followed from the so-called (torsionless) tetrad postulate 
$ \nabla_\mu e_\nu^{a} + \omega_{\mu~b}^{~a} e_\nu^{~b} = 0$,
Eq.~\eqref{eq:Khat2} enables us to obtain 
\begin{equation}
 \int d^4 x 
  \frac{\delta \hK}{\delta \pt (x)} \lie_\beta \pt (x)
  = \int d^4 x \sqrt{\gamma} \delta (\pt - \pt(x)) 
  \left[ \hT^\mu_{~a} \lie_\beta e_\mu^{~a}
   + \hJ^{\mu} \delta_{\lambda} A_{\mu}
   -  \frac{1}{8}C\nu\varepsilon^{\mu\nu\rho\sigma}
   F_{\mu\nu} F_{\rho\sigma} \right]
  \beta',
  \label{eq:Psit}
\end{equation}
where we defined $\beta' \equiv - \beta^\mu n_\mu$ and used 
the operator identity $\hT^{ab} - \hT^{ba} =0$.
By using the identity
\begin{equation}
 n_\alpha \varepsilon^{\mu\nu\rho\sigma} F_{\mu\nu} F_{\rho\sigma}
  = - 4 \varepsilon^{\mu\nu\rho\sigma} n_\nu F_{\rho\sigma} F_{\alpha\mu}.
\end{equation}
the last term in the second line of Eq.~\eqref{eq:Psit} can be further 
simplified as 
\begin{equation}
 \begin{split}
  \int d^4 x \sqrt{\gamma} \delta (\pt - \pt(x)) 
  \left[ \frac{1}{8}C \nu \beta^\alpha n_\alpha \varepsilon^{\mu\nu\rho\sigma}
  F_{\mu\nu} F_{\rho\sigma} \right]
  &= -
  \int d^4 x \sqrt{\gamma} \delta (\pt - \pt(x)) 
  \left[ \frac{1}{2}C \nu \beta^\alpha n_\nu \varepsilon^{\mu\nu\rho\sigma}
  F_{\rho\sigma} F_{\alpha\mu} \right]
  \\
  &= -
  \int d^4 x \sqrt{\gamma} \delta (\pt - \pt(x)) 
  C \nu B^\mu \delta_\lambda A_\mu .
  \\
 \end{split}
\end{equation}
Here we defined the four-magnetic field as $B^\mu \equiv \tilF^{\mu\nu} n_\nu
=  \varepsilon^{\mu\nu\rho\sigma} n_\nu F_{\rho\sigma}/2$, 
and neglected the surface term accompanied by the integration by parts. 
We thus obtain the following compact result:
\begin{equation}
 \int d^4 x \frac{\delta \hK}{\delta \pt (x)} \lie_\beta \pt (x)
  = \int d^4 x \beta' \sqrt{\gamma} \delta (\pt - \pt(x)) 
  \left[ \hT^\mu_{~a} \lie_\beta e_\mu^{~a}
   + \big( \hJ^{\mu} -  C \beta'^{-1}\nu B^\mu \big) \delta_{\lambda} A_{\mu}
  \right].
  \label{eq:Psit2}
\end{equation}
Equipped with this formula together with $\lie_\beta \beta^\mu = 0$, and
$\lie_\beta \nu 
= \lie_\beta (\nu-\beta\cdot A) + \beta^\mu \lie_\beta A_\mu
= \beta^\mu \delta_\lambda A_\mu$,
we are now ready to express $\delta^{\mathrm{para}}_\lambda \hK$ 
in Eq.~\eqref{eq:delhKpara} by the use of the variation of the vierbein and gauge field:
\begin{equation}
 \delta_\lambda^{\mathrm{para}} \hK
  = \int d^3 x
  \left[
   \left(
    \beta' \sqrt{\gamma} \hT^\mu_{~a}
    + \frac{\delta \hK}{\delta e_\mu^{~a}}
   \right)
   \lie_\beta e_\mu^{~a}
   + \left(
      \beta' \sqrt{\gamma} 
      \left[ \hJ^{\mu} -  C \beta'^{-1} \nu B^\mu \right]
      +\frac{\delta \hK}{\delta \nu} \beta^\mu 
      + \frac{\delta \hK}{\delta A_\mu}
	 \right)
   \delta_{\lambda} A_{\mu}
	  \right].
\end{equation}
Let us then take the average of this operator identity over 
the local Gibbs distribution $\hrhoLG[\lambda;t]$.
In the absence of the quantum anomaly, 
we can simply replace the averaged variation of $\hK$ with
the variation of the Massieu-Planck functional: 
$\averageLG{\delta \hK/ \delta j}_t = - \delta \Psi/\delta j$.
Nevertheless, since we are considering the system with the chiral anomaly,
we need to be careful when we take the variation of the charge density 
coupled to the local chemical potential.
Using the relation $\partial (e\hJ^0)/\partial A_\mu = \sqrt{\gamma} C B^\mu$
resulting from the covariant anomaly, we can show
\begin{equation}
 \frac{\delta \hK}{\delta A_\mu}
  = e \hJ^0 \beta^\mu
  + \sqrt{\gamma} \nu C B^\mu
  - \beta' \sqrt{\gamma} \frac{\partial \hLcal}{\partial A_\mu}.
\end{equation}
We can then identify the local Gibbs average of the last term in this 
equation as the covariant current in thermal spacetime, which results in 
the sum of the consistent current and the Bardeen-Zumino current 
composed of $\tilA_\mu$:
\begin{equation}
 \beta' \sqrt{\gamma} 
  \AverageLG{\frac{\partial \hLcal}{\partial A_\mu}}_t
  = \Ncal \int \Dcal \xi \Dcal \xi^\dag 
  e^{\tilScal [\xi,\xi^\dag;\tilA,\tile]}
  \frac{\delta \tilScal}{\delta \tilA_\mu}
  = \frac{\delta \Psi}{\delta \tilA_\mu} 
  - \frac{C}{6} \tilvare^{\mu\nu\rho\sigma} \tilA_\nu \Fcal_{\rho\sigma},
\end{equation}
where $\Ncal$ denotes a normalization constant, and we introduced 
a field strength tensor in thermal spacetime 
$\Fcal_{\mu\nu} \equiv 
\tilpartial_\mu \tilA_\nu - \tilpartial_\nu \tilA_\mu$
together with the totally antisymmetric tensor 
$\tilvare^{\mu\nu\rho\sigma} \equiv N (\beta_0 /\beta') \varepsilon^{\mu\nu\rho\sigma}$.
Using this together with 
$ \averageLG{\delta \hK/\delta e_\mu^{~a}}_t 
= - \delta \Psi/\delta e_\mu^{~a}$, we eventually obtain 
the following identity:
\begin{equation}
  \averageLG{\delta_\lambda^{\mathrm{para}} \hK}_t
  = \int d^3 x
  \Bigg[
  \left(
  \beta' \sqrt{\gamma} \averageLG{\hT^\mu_{~a}}_t
  - \frac{\delta \Psi}{\delta e_\mu^{~a}}
  \right)
  \lie_\beta e_\mu^{~a} 
  + \left(
  \beta' \sqrt{\gamma} 
  \averageLG{\hJ^{\mu}}_t 
  - \frac{\delta \Psi}{\delta \tilA_\mu}
  + \frac{C}{6} \tilvare^{\mu\nu\rho\sigma} \tilA_\nu \Fcal_{\rho\sigma}
  \right)
  \delta_{\lambda} A_{\mu}
  \Bigg].
\end{equation}
Therefore, noting that that this identity holds for an arbitrary variation of 
the background vierbein and gauge field,
the identity $\averageLG{\delta_\lambda^{\mathrm{para}} \hK}_t = 0$ provides 
the variational formula for the Masseiu-Planck functional
\begin{align}
 \averageLG{\hT^\mu_{~a}(t,\bx)}_\pt 
 &= \frac{1 }{\beta' \sqrt{\gamma}} \frac{\delta\Psi[\lambda,j,t]}{\delta
 e_\mu^{~a}(x)} ,
 \label{eq:VariationEM}
 \\
 \averageLG{\hJ^{\mu}(t,\bx)}_\pt 
 &= \frac{1}{\beta' \sqrt{\gamma}} 
 \frac{\delta \Psi[\lambda,j,t]}{\delta \tilA_{\mu}(x)} 
 - \frac{C}{6} \tilvare^{\mu\nu\rho\sigma} \tilA_\nu \Fcal_{\rho\sigma}.
 \label{eq:VariationJ}
\end{align}
We thus conclude that the average values of any conserved current 
operator over local thermal equilibrium is fully captured by the
single (local thermodynamic) functional known as the Masseiu-Planck functional. 
It is worth pointing out that because we deal with the average 
of the covariant current $\averageLG{\hJ(x)}_t$, we have the last term
in Eq.~\eqref{eq:VariationJ} analogous to 
the Bardeen-Zumino current~\cite{Bardeen:1984pm} (See also Refs.~\cite{Banerjee:2012iz,Haehl:2015pja,Landsteiner:2016led}).
In summary, we can identify the Massieu-Planck functional $\Psi[\lambda,j;t]$
as a generating functional for a (nondissipative) local equilibrium part 
of hydrodynamics, or $\averageLG{\hcurrent^\mu_{~a}(t,\bx)}_t$.

Before moving to the path-integral formula for the Massieu-Planck functional,
we put a short comment on the useful ``gauge and coordinate choice'', 
which we call \textit{hydrostatic gauge}.
Since we have a freedom to choose the local time-direction and 
time-component of the external gauge field, we can employ 
the hydrostatic gauge fixing condition
\begin{equation}
 t^\mu (x) = \beta^\mu(x)/\beta_0, \quad
  t^\mu (x) A_\mu (x) = \nu (x) /\beta_0,
\end{equation}
with a constant reference temperature $\beta_0$.
In this special choice of the gauge, the above transformation does not 
induce the gauge transformation because 
$\theta = \epsilon(\nu - \beta \cdot A) = 0$, and furthermore,
thanks to the refined choice of our local time-direction, 
the fluid looks like entirely at rest.
This is the origin of the name hydro\textit{static}.
Nevertheless, note that this does not means the system is in 
a stationary hydrostatic state since we do not assume 
$\beta^\mu$ is a killing vector: $\lie_\beta g_{\mu\nu} \neq 0$.
The main reason why the hydrostatic gauge gives the most useful gauge 
is that we can equate the background field in original (real) spacetime 
with that in (imaginary) thermal spacetime:
$e_\mu^{~a}|_{\HS} = \tile_\mu^{~a}$ and $A_\mu |_{\HS} = \tilA_\mu$.
As a result, the above variational formula results in
\eqref{eq:VariationEM}-\eqref{eq:VariationJ} as 
\begin{align}
 \averageLG{\hT^\mu_{~a}(t,\bx)}_\pt 
 &= \left.
 \frac{1}{\beta_0 e} 
 \frac{\delta\Psi[\lambda,j;t]}{\delta e_\mu^{~a}(x)} 
 \right|_{\HS}
 ,
 \label{eq:HSVariationEM}
 \\
 \averageLG{\hJ^{\mu}(t,\bx)}_\pt 
 &= \left. 
 \frac{1}{\beta_0 e} 
 \frac{\delta \Psi[\lambda,j;t]}{\delta A_{\mu}(x)} 
 \right|_{\HS}
 - \frac{C}{6} 
 \left.
 \varepsilon^{\mu\nu\rho\sigma} A_\nu F_{\rho\sigma}
 \right|_{\HS},
 \label{eq:HSVariationJ}
\end{align}
which enable us to regard the Massieu-Planck functional as a 
usual generating functional.

\subsection{Anomaly matching for local thermodynamic functional}

Based on the obtained formulae, we now discuss the anomaly-induced 
transport from the point of view of the anomaly matching for the 
Massieu-Planck functional.

Before moving to the anomaly-induced transport, let us briefly see 
how we can derive the constitutive relation for a perfect fluid.
Employing the simplest power counting scheme 
$\lambda = O(\nabla^0),~j = O(\nabla^0)$, 
we perform the derivative expansion of the Massieu-Planck function as follows:
\begin{equation}
 \Psi [\lambda,j;t] 
  = \Psi^{(0)} [\lambda,j;t] + \Psi^{(1)} [\lambda,j;t] 
  + O (\nabla^2),
\end{equation}
where the superscript represents the number of spatial derivatives 
acting on parameters $\lambda$ and $j$.
Then, the symmetry argument reviewed in the previous subsection tells us 
that we cannot use the Kaluza-Klein and $U(1)$ 
gauge fields in the leading-order derivative expansion. 
As a result, the general form of the leading-order Massieu-Planck functional
$\Psi^{(0)} [\lambda,j;t]$ is expressed as 
\begin{equation}
 \Psi^{(0)} [\lambda,j;t]
  = \int_0^{\beta_0} d \tau d^3 x \tile 
  p (\beta,\nu)
  = \int d^3 x \beta' \sqrt{\gamma} p (\beta,\nu),
\end{equation}
where $p(\beta,\nu)$ is a certain function depending on $\beta$ and $\nu$.
By taking the variation with respect to the vierbein and gauge field,
we are able to obtain the leading-order constitutive relation as
\begin{equation}
 \averageLG{\hT^{\mu\nu}(t,\bx)}_{(0)} 
  = (e+p) u^\mu u^\nu + p g^{\mu\nu} + O(\nabla^1), \quad
 \averageLG{\hJ^{\mu}(t,\bx)}_{(0)}
  = n u^\mu + O(\nabla^1).
\end{equation}
This is nothing but the constitutive relation for the perfect fluid 
with $e$, $n$, $p$ being the energy density, charge density, and fluid pressure,
respectively.

Then, the next problem is to specify the first-order derivative correction of 
the Massieu-Planck functional $\Psi^{(1)}[\lambda,j;t]$, which is present (absent) in the absence (presence) of the parity symmetry.
Since our system is composed of the right-handed Weyl fermion, and thus, 
there is no parity symmetry, the first-order correction is not prohibited.
In this case, two (anomalous) gauge symmetries again plays a central role 
to extract information on the anomaly-induced transport contained in 
$\Psi^{(1)}[\lambda,j;t]$.
In the following, after giving a bottom up view relying on the one-loop result 
in the previous section, we switch to a top down view of the anomaly matching, 
from which we can derive the anomaly-induced transport beyond the one-loop 
level.

\subsubsection{Chiral anomaly in thermal spacetime}
At one-loop level, we have already derived the anomaly-induced transport
given in Eq.~\eqref{eq:CME-CVE2}.
On the other hand, we also have the variational formula \eqref{eq:VariationJ}
in a general gauge, or \eqref{eq:HSVariationJ} in the hydrostatic gauge.
Let us take the hydrostatic gauge.
Then, the combination of the above results enables us to obtain the 
following functional differential equation for $\Psi^{(1)}$:
\begin{equation}
 \frac{1}{\beta_0} 
  \left. 
   \frac{\delta \Psi_{\ano}^{(1)}[\lambda,j;t]}{\delta A_i (x)} 
  \right|_{\HS}
  +\frac{\mu}{12 \pi^2} B^i -\frac{1}{12\pi^{2}}\varepsilon^{0ijk}A_{k}\partial_{j}\mu
  = \frac{\mu}{4\pi^2} B^i 
  + \left( \frac{\mu^2}{4\pi^2} 
  + \frac{T^2}{12} \right) \omega^i ,
\end{equation}
where we take the flat limit and assume global thermal equilibrium 
with a constant temperature $\beta_0$ in the variational formula.
This equation can be easily solved as
\begin{equation}
 \begin{split}
  \left. \Psi_{\ano}^{(1)} [\lambda,j;t] \right|_{\HS}^{\mathrm{eq}}
  &= \frac{\beta_0}{12\pi^2} \int d^3 x 
  \mu A_i B^i 
  + \beta_0 \int d^3 x 
  \left( \frac{1}{4\pi^2} \mu^2 + \frac{1}{12} T^2  \right)
  A_i \omega^i 
  \\
  &= \frac{\beta_0}{12\pi^2} \int d^3 x 
   \varepsilon^{0ijk} 
  \mu A_i \partial_j A_k
  + \frac{\beta_0}{2} \int d^3 x 
  \varepsilon^{0ijk}
  \left( \frac{1}{4\pi^2} \mu^2 + \frac{1}{12} T^2  \right)
  A_i \partial_j u_k
 \end{split}
 \label{eq:Psi1flat}
\end{equation}
up to irrelevant constants.
On the other hand, we have already clarified that the Massieu-Planck 
functional need to respect both $U(1)$ and Kaluza-Klein gauge invariance.
This constraint then enables us to guess the full result on $\Psi^{(1)}$ 
for general local thermal equilibrium though Eq.~\eqref{eq:Psi1flat} 
is obtained by matching with the one-loop result  
for linear perturbations on the top of global thermal equilibrium.
By using the $U(1)$ and Kaluza-Klein gauge covariant 
quantities---$\tilA_i'$ and $a_i$, respectively---together with 
$\tilA_0 = e^\sigma \mu$, we specify the first-order derivative correction as 
\begin{equation}
 \Psi_{\ano}^{(1)} [\lambda,j;t]
  =  \frac{C\beta_0}{3} \int d^3 x e \varepsilon^{0ijk} 
  \tilA_0 \tilA_i' \partial_j \tilA_k'
  + \frac{C\beta_0}{6} \int d^3 x e \varepsilon^{0ijk} 
  \tilA_0^2 \tilA_i' \partial_j a_k,
  - \frac{C_1}{2\beta_0} \int d^3 x e \varepsilon^{0ijk} 
  \tilA_i' \partial_j a_k,
 \label{eq:Psi1}
\end{equation} 
with $C_1 \equiv 1/12$. 
Note that $\tilA_0$ and $\tilA_i'$ defined in 
Eqs.~\eqref{eq:Thermalbg} and \eqref{eq:KKinv}
are manifestly Kaluza-Klein gauge invariant quantities.

Let us then confirm the consistency for this result 
based on the anomaly matching for the Massieu-Planck functional itself.
For that purpose, we consider the time-independent gauge transformation 
given by
$\delta_\theta \tilA_0 = 0,~ \delta_\theta \tilA_i = \partial_i \theta (\bx)$.
Under this gauge transformation, the Fujikawa method~\cite{Fujikawa-Suzuki} 
says that the anomalous shift of the Massieu-Planck functional is given 
by the consistent anomaly:
\begin{equation}
  \delta_\theta \Psi [\lambda,j;t]
  = - \frac{C \beta_0}{3} \int d^3 x 
  \theta e \varepsilon^{0ijk} \partial_i \tilA_0 \partial_j \tilA_k.
\end{equation}
On the other hand, one can directly show that 
the first two term of $\Psi_{\ano}^{(1)} [\lambda,j;t]$ in Eq.~\eqref{eq:Psi1}
correctly reproduces this anomalous shift as
\begin{align}
 \delta_\theta \Psi_{\ano}^{(1)} [\lambda,j;t]
 &= \frac{C\beta_0}{3} \int d^3 x e \varepsilon^{0ijk}
 \tilA_0 \partial_i \theta \partial_j \tilA_k'
 + \frac{C\beta_0}{6} \int d^3 x e \varepsilon^{0ijk}
 \tilA_0^2 \partial_i \theta \partial_j a_k
 \nonumber
 \\
 &= - \frac{C\beta_0}{3} \int d^3 x \theta e \varepsilon^{0ijk} 
 \partial_i\tilA_0 \partial_j (\tilA_k - \tilA_0 a_k)
 - \frac{C\beta_0}{3} \int d^3 x \theta
 e \varepsilon^{0ijk} \tilA_0 \partial_i \tilA_0 \partial_j a_k
 + (\mathrm{surface~terms})
 \nonumber
  \\
 &= - \frac{C\beta_0}{3} \int d^3 x 
 \theta \varepsilon^{0ijk} \partial_i\tilA_0 \partial_j \tilA_k
 + (\mathrm{surface~terms}).
\end{align}
Therefore, we see that the anomalous transport coefficients $C$ 
proportional to the chemical potential $\mu$ is indeed related to the anomaly 
coefficient attached to the Weyl fermion. 

Nevertheless, the last term in Eq.~\eqref{eq:Psi1}, 
which brings about the chiral vortical effect proportional to $T^2$, 
is not restricted by the chiral anomaly.
From the symmetry point of view, this is just because 
the last term in Eq.~\eqref{eq:Psi1} remains \textit{invariant} under 
the $U(1)$ gauge transformation.
This corresponds the fact that the entropy production argument with 
chiral anomaly leads to the existence of both chiral magnetic and vortical effect~\cite{Son:2009tf}, in which only the anomalous transport coefficients proportional to the chemical potential are determined.
Then, the natural question is 
\textit{''Does the chiral vortical effect proportional to $T^2$ have any 
relation with the quantum anomaly?''}

\subsubsection{Global anomaly for Kaluza-Klein gauge transformation}
It was pointed out the $T^{2}$ term of the chiral vortical coefficient is related to the gravitational contribution to the chiral anomaly~\cite{Landsteiner:2011cp}. However, unlike the chiral magnetic coefficient discussed in this section, it is not clear that how the chiral vortical effect relates to the $\varepsilon^{\mu\nu\rho\sigma} R^{\alpha}_{~\beta\mu\nu} R^{\beta}_{~\alpha\rho\sigma}$, because the number of derivative in $\varepsilon^{\mu\nu\rho\sigma} R^{\alpha}_{~\beta\mu\nu} R^{\beta}_{~\alpha\rho\sigma}$ is higher than that in $\varepsilon^{\mu\nu\rho\sigma} F_{\mu\nu} F_{\rho\sigma}$. In other words, $\varepsilon^{\mu\nu\rho\sigma} R^{\alpha}_{~\beta\mu\nu} R^{\beta}_{~\alpha\rho\sigma}$ does not directly contribute to the first order hydrodynamics. 
An alternative explanation of $T^{2}$ term is that the chiral vortical coefficient is related to a global anomaly~\cite{Golkar:2015oxw,Chowdhury:2016cmh,Nakai:2016lle}. Here, we show the relation between the global anomaly and chiral vortical effect. 

As a warm up exercise, let us first consider the global anomaly attached to the
Weyl fermion in  $1+1$ dimensions, which possesses the chiral anomaly given by
\begin{equation}
 \partial_\mu \hJ^\mu = - \frac{1}{2} C_{2D} \varepsilon^{\mu\nu} F_{\mu\nu}
  \with 
  C_{2D} \equiv \frac{1}{2\pi},
\end{equation}
where $\hJ^\mu$ again denotes the covariant current in $1+1$ dimensional system.
In this case, there are no chiral magnetic and vortical effects because there is no transverse direction, and thus, no magnetic field and vorticity.
However, there exist nonvanishing  $\langle \hJ^{z}\rangle$ and 
$\langle \hT^0_{~z}\rangle$ caused by chiral and global anomalies.
The direct calculation at equilibrium shows 
\begin{equation}
 \begin{split}
  \averageG{\hT^0_{~z}}
  &= \int_{0}^{\infty} \frac{dp_{z}}{2\pi}p_{z}
  \big[  n_{F}(|p_{z}|-\mu) +n_{F}(|p_{z}|+\mu) \big]
  = \frac{\mu^{2}}{4\pi} + \frac{\pi}{12}T^{2},
  \\
  \averageG{\hJ^{z}}
  &= \int_{0}^{\infty} \frac{dp_{z}}{2\pi} \frac{p_{z}}{|p_{z}|}
  \big[ n_{F}(|p_{z}|-\mu) -n_{F}(|p_{z}|+\mu) \big]
  =\frac{\mu}{2\pi}.
 \end{split}
\end{equation}
On the other hand, the same procedure given above leads to 
the variational formula in $(1+1)$ dimensions:
\begin{align}
 \begin{split}
 \averageLG{\hT^\mu_{~a}}_\pt 
 &= \frac{1}{\beta' \sqrt{\gamma}} \frac{\delta\Psi[\pt;\lambda]}{\delta
 e_\mu^{~a}(x)} ,
  \\
 \averageLG{\hJ^{\mu}}_\pt 
 &= \frac{1}{\beta' \sqrt{\gamma}} 
  \frac{\delta \Psi[\pt;\lambda]}{\delta \tilA_{\mu}(x)} 
  -\frac{1}{2}C_{2D} \tilvare^{\mu\nu} \tilA_{\nu},
 \end{split}
  \label{eq:1+1Variation}
\end{align}
where $\tilvare^{\mu\nu}=N(\beta_{0}/\beta') \varepsilon^{\mu\nu}$.
Then, the matching condition for the momentum density and current results in
\begin{align}
 &{ \frac{1}{\beta_0 }
 \frac{\delta \Psi_\ano}{\delta e_0^{~z}} }
 = - \frac{1}{\beta_0}\frac{\delta \Psi_\ano}{\delta a_{z}}
 = \frac{C_{2D}}{2}\mu^{2} + \pi C_{1}T^{2} 
 ,
 \label{eq:T0z}
 \\
 &{\frac{1}{\beta_0}} \frac{\delta \Psi_\ano}{\delta A_{z}} 
  + \frac{C_{2D}}{2}  \mu
 = C_{2D} \mu \label{eq:jz}.
\end{align} 
Solving Eqs.~\eqref{eq:T0z} and \eqref{eq:jz}, we find 
\begin{equation}
\Psi_\ano 
 = \frac{C_{2D} \beta_0}{2}
 \int dz \tilA_0
 \tilA_{z}' 
 - \pi \frac{C_1}{\beta_0} \int dz a_{z}.
\end{equation}
This gives the anomalous part of the Masseiu-Planck functional.
In order to detect anomalies,
we compactify the spatial direction with the length $L$. 
Here we will show $\Psi_\ano$ has two types of anomalies. 
One is the chiral anomaly:
Under $U(1)$ gauge transformation $\tilA_z \to \tilA_z + \partial_z \theta (z)$,
the anomalous shift of $\Psi$ arises:
\begin{equation}
 \begin{split}
  \delta_\theta \Psi_\ano 
  = - {\frac{C_{2D}\beta_0}{2}} \int dz \theta \partial_{z} \tilA_0,
 \end{split}
\end{equation}
which correctly reproduces the consistent anomaly in thermal spacetime.
The other is the global anomaly associated with the Kaluza-Klein gauge 
transformation:
\begin{equation}
 \begin{cases}
  \tilt \to \tilt + \chi(z), \\
  a_{z} \to a_{z} - \partial_{z}\chi(z),
 \end{cases}
\end{equation}
where ${\tilA_{z}'}$ remains invariant. 
Under this transformation, $\Psi_\ano$ also acquires the anomalous shift
given by
\begin{equation}
 \begin{split}
  \delta_\chi \Psi_\ano 
  = \pi {\frac{C_1}{\beta_0}} 
  \int dz \partial_{z}\chi(z),
 \end{split}
\end{equation}
which is just a boundary term, so that $\Psi_\ano $ is invariant
under local transformation with $\chi(0)=\chi(L)$. 
However, if we consider global transformation, 
$\chi(z) = - 2 i\beta_0 z /L$, 
which corresponds to the imaginary time shift 
$\tau\to \tau+ 2 z \beta_0/L$ that keep the boundary condition, we have an additional phase
\begin{equation}
\begin{split}
\Psi_\ano \to  \Psi_\ano -2\pi iC_{1},
\end{split}
\end{equation}
which can be understood as the global anomaly associated with the large diffeomorphism.
This anomalous phase is related to the three dimensional gravitational Chern-Simons term through the anomaly inflow mechanism, which is also related to the gravitational contribution to chiral anomaly in $3+1$ dimensions~\cite{Witten:1983tw,Witten:1985xe}.

This argument can be generalized to higher dimensions. In $(3+1)$ dimensions, $\Psi_\ano$ is given in Eq.~\eqref{eq:Psi1}. 
In order to detect the global anomaly, we compactify the space to $S^{1}\times S^{2}$, where we choose $z$ as the coordinate on $S^{1}$.
Under the large diffeomorphism, $\tau\to \tau+ 2 z\beta_0/L$,
the term contributing to the $T^{2}$ part of chiral vortical effect transforms as
\begin{equation}
\begin{split}
 \Psi_\ano \to 
 \Psi_\ano  -2\pi iC_1 \int \frac{d^2x}{2\pi} e \varepsilon^{0ijz} 
 \partial_i \tilA_j'.
\end{split}
\end{equation}
This is the global mixed anomaly between $U(1)$ gauge and large diffeomorphism.
Therefore, we see that the chiral vortical coefficient proportional to $T^{2}$, 
which is nothing but $C_{1}$, is related to the mixed global anomaly.

\section{Summary and discussion}
\label{sec:Summary}
In this paper, we have discussed two approaches to derive the anomaly-induced 
transport phenomena for the system composed of a Weyl fermion: 
perturbative evaluation of the chiral magnetic/vortical conductivity 
with the help of the (equilibrium) linear response theory, 
and the nonperturbative determination of anomalous parts of 
the local thermodynamic functional on the basis of the anomaly matching.
Both derivations are based on the imaginary-time formalism of the 
quantum field theory, and we have seen that the obtained anomalous constitutive 
relations correctly describe the chiral magnetic/vortical effect.
Although it is not so clear in the first derivation, 
the second derivation shows that the chiral magnetic/vortical effect 
results from the first-order derivative corrections of the local thermodynamic 
functional, and thus, they are clearly nondissipative in nature.
This is perfectly consistent with the known result obtained from the 
hydrostatic partition function method~\cite{Banerjee:2012iz,Jensen:2012jh,Jensen:2012jy,Jensen:2012kj,Banerjee:2012cr,Jensen:2013vta,Jensen:2013kka,Jensen:2013rga,Haehl:2015pja,Golkar:2015oxw,Chowdhury:2016cmh}, 
and we rigorously clarify why that method works well.
This local equilibrium part of the constitutive relation
also complete the application of Zubarev's nonequilibrium statistical operator 
method to derive the hydrodynamic equation for the 
parity-violating (anomalous) fluid.

There are several interesting questions related to the current work.
It has been already pointed out that the coefficient in front of the $T^2$-term 
of the chiral vortical effect will be renormalized in the presence of 
dynamical gauge fields such as the gluon in the QCD plasma~\cite{Golkar:2012kb}.
It may be interesting to examine which part of the anomaly matching argument
associated with the large diffeomorphism (Kaluza-Klein gauge) transformation
should be modified due to the existence of the dynamical gauge field.
Another important issue associated with the inclusion of 
dynamical electromagnetic field is its dynamics.
When we consider the dynamics of the electromagnetic field rather than 
treating it as the background one, we encounter with several interesting 
phenomena such as the chiral plasma instability~\cite{Boyarsky:2011uy,Tashiro:2012mf,Akamatsu:2013pjd,Akamatsu:2014yza,Manuel:2015zpa}, and mixing of 
some hydrodynamic modes (chiral magnetic wave) to be the massive collective 
excitation (chiral plasmon)~\cite{Kharzeev:2010gd,Gorbar:2016ygi,Gorbar:2016sey,Rybalka:2018uzh}.
It is desirable to systematically describe them 
based on the generalization of magnetohydrodynamics
for the chiral plasma by formulating chiral magnetohydrodynamics.
Chiral magnetohydrodynamics is just recently formulated based on
e.g., the phenomenological entropy-current analysis~\cite{Hattori:2017usa} 
(See also Refs.~\cite{Boyarsky:2015faa,Gorbar:2016qfh,Yamamoto:2016xtu,Giovannini:2016whv,Rogachevskii:2017uyc}),
but less is clarified from the underlying quantum field theory.
Combined with the recent development of the magnetohydrodynamics itself 
from the field theoretical viewpoint~\cite{Huang:2011dc,Grozdanov:2016tdf,Hernandez:2017mch,Glorioso:2018kcp,Armas:2018zbe}, it may be interesting to formulate 
chiral magnetohydrodynamics based on the Zubarev's nonequilibrium 
statistical operator method equipped with the path-integral formula 
for the local thermodynamic functional reviewed in this paper.

%%%%%%%%%%%%%%%%%%%%%%%%%%%%%%%%%%%%%%%%%%
\vspace{6pt} 

%%%%%%%%%%%%%%%%%%%%%%%%%%%%%%%%%%%%%%%%%%
%% optional
%\supplementary{The following are available online at \linksupplementary{s1}, Figure S1: title, Table S1: title, Video S1: title.}


%%%%%%%%%%%%%%%%%%%%%%%%%%%%%%%%%%%%%%%%%%
% \authorcontributions{For research articles with several authors, a short paragraph specifying their individual contributions must be provided. The following statements should be used “conceptualization, X.X. and Y.Y.; methodology, X.X.; software, X.X.; validation, X.X., Y.Y. and Z.Z.; formal analysis, X.X.; investigation, X.X.; resources, X.X.; data curation, X.X.; writing—original draft preparation, X.X.; writing—review and editing, X.X.; visualization, X.X.; supervision, X.X.; project administration, X.X.; funding acquisition, Y.Y.”, please turn to the  \href{http://img.mdpi.org/data/contributor-role-instruction.pdf}{CRediT taxonomy} for the term explanation. Authorship must be limited to those who have contributed substantially to the work reported.}

%%%%%%%%%%%%%%%%%%%%%%%%%%%%%%%%%%%%%%%%%%
\funding{This research was funded by Japan Society of Promotion of Science (JSPS) Grant-in-Aid for Scientific Research grant number JP16J02240, 16K17716, 17H06462, and 18H01211.}
%\funding{Please add: ``This research received no external funding'' or ``This research was funded by NAME OF FUNDER grant number XXX.'' and  and ``The APC was funded by XXX''. Check carefully that the details given are accurate and use the standard spelling of funding agency names at \url{https://search.crossref.org/funding}, any errors may affect your future funding.}

%%%%%%%%%%%%%%%%%%%%%%%%%%%%%%%%%%%%%%%%%%
\acknowledgments{
M.~H. was supported by the Special Postdoctoral Researchers Program at RIKEN.
This work was partially supported by the RIKEN iTHES/iTHEMS Program, 
in particular, iTHEMS STAMP working group. 
% In this section you can acknowledge any support given which is not covered by the author contribution or funding sections. This may include administrative and technical support, or donations in kind (e.g., materials used for experiments).
}

%%%%%%%%%%%%%%%%%%%%%%%%%%%%%%%%%%%%%%%%%%
% \conflictsofinterest{Declare conflicts of interest or state ``The authors declare no conflict of interest.'' Authors must identify and declare any personal circumstances or interest that may be perceived as inappropriately influencing the representation or interpretation of reported research results. Any role of the funders in the design of the study; in the collection, analyses or interpretation of data; in the writing of the manuscript, or in the decision to publish the results must be declared in this section. If there is no role, please state ``The funders had no role in the design of the study; in the collection, analyses, or interpretation of data; in the writing of the manuscript, or in the decision to publish the results''.} 

%%%%%%%%%%%%%%%%%%%%%%%%%%%%%%%%%%%%%%%%%%
%% optional
% \abbreviations{The following abbreviations are used in this manuscript:\\

% \noindent 
% \begin{tabular}{@{}ll}
% MDPI & Multidisciplinary Digital Publishing Institute\\
% DOAJ & Directory of open access journals\\
% TLA & Three letter acronym\\
% LD & linear dichroism
% \end{tabular}}

%%%%%%%%%%%%%%%%%%%%%%%%%%%%%%%%%%%%%%%%%%
%% optional
\appendixtitles{no} %Leave argument "no" if all appendix headings stay EMPTY (then no dot is printed after "Appendix A"). If the appendix sections contain a heading then change the argument to "yes".
\appendixsections{multiple} %Leave argument "multiple" if there are multiple sections. Then a counter is printed ("Appendix A"). If there is only one appendix section then change the argument to "one" and no counter is printed ("Appendix").
\appendix


%=====================================
% References, variant A: internal bibliography
%=====================================
\reftitle{References}

\bibliography{Anomalous}

% \bibitem[Author1(year)]{ref-journal}
% Author1, T. The title of the cited article. {\em Journal Abbreviation} {\bf 2008}, {\em 10}, 142-149, doi:xxxxx.
% % Reference 2
% \bibitem[Author2(year)]{ref-book}
% Author2, L. The title of the cited contribution. In {\em The Book Title}; Editor1, F., Editor2, A., Eds.; Publishing House: City, Country, 2007; pp. 32-58, ISBN.
% \end{thebibliography}

% The following MDPI journals use author-date citation: Arts, Econometrics, Economies, Genealogy, Humanities, IJFS, JRFM, Laws, Religions, Risks, Social Sciences. For those journals, please follow the formatting guidelines on http://www.mdpi.com/authors/references
% To cite two works by the same author: \citeauthor{ref-journal-1a} (\citeyear{ref-journal-1a}, \citeyear{ref-journal-1b}). This produces: Whittaker (1967, 1975)
% To cite two works by the same author with specific pages: \citeauthor{ref-journal-3a} (\citeyear{ref-journal-3a}, p. 328; \citeyear{ref-journal-3b}, p.475). This produces: Wong (1999, p. 328; 2000, p. 475)

%=====================================
% References, variant B: external bibliography
%=====================================
%\externalbibliography{yes}
%\bibliography{your_external_BibTeX_file}

%%%%%%%%%%%%%%%%%%%%%%%%%%%%%%%%%%%%%%%%%%
%% optional
% \sampleavailability{Samples of the compounds ...... are available from the authors.}

%% for journal Sci
%\reviewreports{\\
%Reviewer 1 comments and authors’ response\\
%Reviewer 2 comments and authors’ response\\
%Reviewer 3 comments and authors’ response
%}

%%%%%%%%%%%%%%%%%%%%%%%%%%%%%%%%%%%%%%%%%%
\end{document}